\documentclass[12pt]{article}%
\usepackage{amsmath,amssymb,amsfonts,cite}
\usepackage{graphicx,epsfig}
\usepackage{color}
\usepackage{amsmath}
\usepackage{amsfonts}
\usepackage{amssymb}
\usepackage{float}
\usepackage{cancel}
\usepackage{caption}
\usepackage{graphicx}
\usepackage{hyperref}%
\numberwithin{equation}{section} 
\begin{document}
\begin{titlepage}
\begin{flushright}
\end{flushright}

\begin{center}
{\huge\bf SUSY without the Little Hierarchy}\\

\medskip
\bigskip\color{black}\vspace{0.6cm}
{
{\large\bf Brando Bellazzini$^{a,b}$, Csaba Cs\'aki$^a$,\\[3mm] Antonio Delgado$^c$, and Andreas Weiler$^{a,d}$}
}
\\[7mm]
{\it $^a$ Institute for High Energy Phenomenology,\\ Newman Laboratory of Elementary Particle Physics,\\
Cornell University, Ithaca, NY 14853, USA} \\
\vspace*{0.3cm}
{\it $^b$ INFN, Sezione di Pisa,  Largo Pontecorvo 3, 56127, Pisa,
Italy}
\\
\vspace*{0.3cm}
{\it $^c$ Department of Physics, University of Notre Dame, Indiana 46556, USA} \\
\vspace*{0.3cm}
{\it $^d$ CERN Theory Division, CH-1211 Geneva 23, Switzerland}
\\
\vspace*{0.3cm}
{\it E-mail: \rm{b.bellazzini@cornell.edu, csaki@cornell.edu, antonio.delgado@nd.edu, andreas.weiler@cern.ch}}
\bigskip\bigskip\bigskip

{
\centerline{\large\bf Abstract}
\begin{quote}

We propose a simple gauge extension of the minimal supersymmetric
standard model (MSSM), where the fine tuning in the Higgs mass
parameters is highly reduced.  The Higgs boson is insensitive to
high energies because of supersymmetry and also because it is a
pseudo-Goldstone boson of a global symmetry (referred to as ``double
protection'' or ``super-little Higgs'' mechanism). A large shift in
the Higgs quartic self coupling is obtained via a non-decoupling
D-term coming from an extra gauge group, resulting in a Higgs that
can be as heavy as $135$ GeV with a tuning milder than 10\%. With an
appropriate choice of quantum numbers one can achieve that the
additional quartic is generated without a corresponding shift in the
Higgs mass, thus preserving the double protection of the Higgs. The
model predicts the existence of several top-partner fermions one of
which can be as light as $700$ GeV, while the ordinary MSSM states
could be as light as $400$ GeV. In addition to many new particles in
the multi-TeV range there would also be an axion-like state very
weakly coupled to standard model (SM) matter which could be in the
sub-GeV regime, and sterile neutrinos which could be light.

\end{quote}}

\end{center}
\end{titlepage}

\section{Introduction}
\label{intro}

Supersymmetry (SUSY) is a very attractive solution to the hierarchy
problem. The electroweak scale is only logarithmically sensitive to
the cutoff scale $\Lambda$, while it is quadratically sensitive to
the soft SUSY breaking parameters $m_s$. Thus if a SUSY model is to
be natural, then the generic prediction would be that the soft
breaking scale has to be around the electroweak symmetry breaking
(EWSB) scale $v\simeq175$ GeV. However, the lack of discovery of
superpartners at LEP2 and the Tevatron, and the bounds from
electroweak precision tests (EWPT) and flavor physics, frustrate
this natural expectation, reintroducing a tension between these two
scales $m_s$ and $v$\cite{Barbieri:2000gf}. This is usually referred
to as the little hierarchy problem.

The way the little hierarchy problem manifests itself in the context
of the minimal supersymmetric standard model (MSSM) is via a tension
between the lightest physical Higgs mass and the mass of the
Z-boson: in order for the Higgs mass to be above the LEP2 bound of
$115$ GeV, one needs to radiatively enhance the Higgs quartic
self-coupling, which requires a large stop mass $m_{s}\gtrsim$~TeV.
However, the same heavy stop will then yield the dominant radiative
corrections to the quadratic Higgs mass parameter through the
stop/top loops. The natural size of the electroweak scale would then
be at
\begin{equation}
(3m_{s}^{2}/4\pi^{2})\ln\Lambda/m_{s},
\end{equation}
which generically results in a percent level (or worse) fine tuning.

This supersymmetric little hierarchy problem stimulated several
authors to depart from SUSY, and to look for alternatives. In little
Higgs (LH) models \cite{ArkaniHamed:2001nc}(for reviews
see~\cite{Schmaltz:2005ky}), the quadratic sensitivity to the cutoff
scale is removed by same spin particles whose couplings are fixed by
a global approximate symmetry broken at some scale $f$: these models
recast the old idea that the Higgs is naturally light because it
emerges as a pseudo-Goldstone boson of a broken global symmetry
\cite{Kaplan:1983fs}. This global symmetry breaking is usually
linked to a new strongly interacting sector \cite{Agashe:2004rs}.
However, LH models generically generate tree level contributions to
electroweak precision observables so that the scale $f$ has to be in
the multi TeV range to evade the electroweak precision bounds
\cite{Hewett:2002px}. Thus the natural EWSB scale of order
\begin{equation}
(3f^{2}/4\pi^{2})\ln f/v
\end{equation}
in generic LH models is still too large. LH models with T-parity can
accommodate EWPT with a lower $f$ \cite{Cheng:2003ju,Hubisz:2005tx},
but they do not provide a solution to the (big) hierarchy problem
between $f$ and $\Lambda$, unless complicated additional layers of
structures are added~\cite{TparityUVcompl}.

One prominent idea is to combine the broken global symmetry and
supersymmetry, to enforce a ``double protection'' (or ``super-little
Higgs'') mechanism
\cite{Birkedal:2004xi,Chankowski:2004mq,Berezhiani:2005pb,Roy:2005hg,SLH,Bellazzini:2008zy}
 on the EWSB scale whose natural value is then expected to be of the order
\begin{equation}
(3m_{s}^{2}/4\pi^{2})\ln f/m_{s}
\end{equation}
where the cutoff $\Lambda$ of the MSSM   is replaced by the global
symmetry breaking scale  $ f\sim$ few $\times$ TeV. All the good
features of SUSY are retained: it solves the hierarchy problem
between $f$ and $\Lambda$ and satisfies EWPT's. However, for a
completely natural theory all superpartners (including the stop)
should be in the few hundred GeV range, which would generically
result in a Higgs mass that is too light \cite{Bellazzini:2008zy}. Therefore a new
contribution to the Higgs quartic self interaction is needed in
order to raise the Higgs mass above the LEP bound. However, the
source for the new quartic should not spoil the double protection
with an unwanted large correction to the quadratic Higgs coupling.
Existing models have to include a complicated new sector with
several light gauge singlets to achieve this~\cite{Roy:2005hg,SLH}.

In this paper we realize this program via non-decoupling D-terms of
extended gauge symmetries~\cite{Batra:2003nj,Maloney:2004rc}, where
the effective gauge couplings, hence the Higgs quartic, are
enhanced.\footnote{Other known mechanisms for enhancing the Higgs
quartic include the NMSSM~\cite{NMSSM} and fat Higgs~\cite{fathiggs}
type models.} With an appropriate choice of quantum numbers under
the extended gauge group we can ensure that the only non-decoupling
effects are in the form of an additional contribution of the
hypercharge D-term. This will guarantee that the quadratic Higgs
terms remain unaffected, and the double protection mechanism
continues to work. The non-decoupling is driven by a sizable soft
breaking mass term for a field that does not carry SM quantum
numbers (but carries only charges under the extended gauge group).
The presence of this field will modify (via interactions through the
D-terms) the effective Lagrangian for the light fields: the
supersymmetric low energy limit for the usual D-terms of the
unbroken generators is replaced by effective D-terms with enhanced
gauge couplings~\cite{Batra:2003nj,Maloney:2004rc}. This framework
provides both the residual approximate global symmetry and the
enhancement of the tree level Higgs quartic coupling. The explicit
model we discuss here is a supersymmetric version of the
``simplest'' LH model \cite{Schmaltz:2004de} based on an $SU(3)_{W}$
gauge symmetry which extends the weak $SU(2)_{W}$. The charge
assignment of the matter content is anomaly free and generation
universal. The model turns out to be natural over a wide range of
the input parameters, requiring a fine tuning of better than $10\%$.
The two main constraints on how little fine tuning one can get away
with are from the requirements that the Z' is sufficiently heavy to
avoid electroweak precision bounds, and that the Landau pole of the
new U(1) gauge group is separated by several orders of magnitude
from the mass scale of the new particles introduced here. Within the
same region without tuning (and satisfying these bounds) the Higgs
mass can be raised up to about 135 GeV. The MSSM masses can be as
light as around 400 GeV (as low as the direct detection bounds allow
them to be). There will be a large variety of new states beyond the
MSSM in the spectrum. The top partners can be as light as 700 GeV,
while most other ``little partner'' states will be in the multi-TeV
regime, including a $Z^{\prime}$. The model also predicts a light
axion-like state, which is very weakly coupled to SM fields, and
sterile neutrinos.

The paper is organized as follows: in Section~\ref{SSLH} we review
the simplest super-little Higgs model, and explain that without
additional gauge extension the theory is still fine tuned. In
Section~\ref{doping} we further extend the gauge group with an
additional U(1) factor, and show that the mechanism
of~\cite{Batra:2003nj,Maloney:2004rc} can be easily implemented. In
Section~\ref{topsect} we include the matter fields into the theory,
and calculate the leading loop corrections of the full model.
Section~\ref{tuning} will  be devoted to quantify the fine tuning in
the theory and to give some interesting signals for the LHC. We
conclude in Section~\ref{conclusions}.

\section{The simplest super-little Higgs}
\label{SSLH}

Let us start considering the supersymmetric version of the simplest
LH model. The electroweak gauge group $SU(2)_{W}\times U(1)_{y}$ of
the Standard Model (SM) is promoted to a larger group,
$SU(3)_{W}\times U(1)_{x}$, under which the Higgs fields are chiral
superfields\footnote{We work with the normalization
$Y=T^{8}/\sqrt{3}+X$ where $X$, $Y$ and $T^{a}$ are the $U(1)_{x}$,
$U(1)_{y}$ and $SU(3)_{W}$ generators respectively, with $T^{3,8}$
diagonal, and $T^8=\lambda^8/2 =1/\sqrt{3} {\rm diag}
(\frac{1}{2},\frac{1}{2},-1)$. Color $SU(3)_{QCD}$ is left untouched
and is assumed to be understood everywhere. The $SU(3)_{W}\times
U(1)_{x}$ gauge couplings are $g$ and $g_{x}$.}. We use a couple of
Higgs fields, $\Phi_{u,d}=3_{+1/3},\,\bar{3}_{-1/3}$, to break the
gauge group down to the SM at the some high scale $F\simeq10$ TeV to
evade all the experimental constraints on new heavy gauge bosons.
Another copy of Higgs fields,
$\mathcal{H}_{u,d}=3_{+1/3},\,\bar{3}_{-1/3}$, takes vacuum
expectation values (VEVs) at a lower scale, $f\simeq1$ TeV: if the
$\mathcal{H}$ VEVs point in the same direction than $\Phi$ VEVs
there is no EWSB, while a misalignment of the $\langle \Phi \rangle$
and $\langle \mathcal{H}\rangle$ VEVs will lead to EWSB.

As long as the fields $\Phi$ and $\mathcal{H}$ do not communicate
with each other, there is an enlarged $SU(3)_{1}\times SU(3)_{2}$
global symmetry with the diagonal $SU(3)_{W}$ gauged. At the scale
$F$ both $SU(3)_{1}$ and $SU(3)_{W}$ are broken, while the global
$SU(3)_{2}$ acting on $\mathcal{H}$ is still preserved down to the
scale $f$. Thus at low energies we are left with $5$ physical
Goldstones living (mostly) in $\mathcal{H}$. The other $5$
longitudinal Goldstones are eaten by the gauge bosons corresponding
to the broken $SU(3)_{W}$ generators. One of physical Goldstone
bosons is an electroweak singlet $\eta$ which does not play any role in the
following. We will discuss it later in Section~\ref{tuning}.
The other $4$ Goldstones form an $SU(2)_{W}$ doublet with
the quantum numbers of the Higgs boson.

To be more concrete, we consider  the following superpotential for
the Higgs sector
\begin{equation}
\mathcal{W}=\kappa N \left(\Phi_{u}\Phi_{d}-\mu^{2} \right)+\mathcal{W}_{\mathcal{H}}(\mathcal{H}_{u,d})
\end{equation}
with no mixing terms like $\Phi_{u}\mathcal{H}_{d}$  that would
spoil the $SU(3)_{1}\times SU(3)_{2}$ symmetry. Here the singlet
field $N$ and the superpotential $\mathcal{W}$ provide just one of
the possible realizations of  the symmetry breaking pattern we
described above. Taking equal soft terms
$m^{2}_{\Phi_{u}}=m^{2}_{\Phi_{d}}\equiv m^{2}>0$ the scalar
potential will be
\begin{equation}
V=\kappa^{2}|\Phi_{u}\Phi_{d}|^{2}-\kappa^{2}\mu^{2}\left(\Phi_{u}\Phi_{d}+h.c.\right)
  +m^{2}(|\Phi_{u}|^{2}+|\Phi_{d}|^{2})+\kappa^{2}|N|^{2}(|\Phi_{u}|^{2}+|\Phi_{d}|^{2})
\label{VF}
\end{equation}
\begin{equation}
\langle\Phi_{u,d}\rangle=(0,0,F)\,,\qquad \langle|N|\rangle=0\,,
 \qquad F^{2}=\mu^{2}-m^{2}/\kappa^{2}>0\,.
 \label{vevsF}
\end{equation}
We will comment later on the effect of having non-equal VEVs,
$F_{u}^{2}-F_{d}^{2}\approx m^{2}_{\Phi_{u}}-m^{2}_{\Phi_{d}}$.

Integrating out at tree level the heavy modes around the VEVs of $N$
and $\Phi_{u,d}$, we get the effective potential for the light
fields\footnote{Here $H_{u,d}$ and $S_{u,d}$ are $SU(2)_{W}\times U(1)_{y}$
doublets and singlets respectively.} $\mathcal{H}_{u,d}$
\begin{equation}
\mathcal{H}_{u}^{T}=(H_{u}^{T},S_{u})\qquad \mathcal{H}_{d}=(H_{d},S_{d})
\end{equation}
 It is given by  the
$SU(3)_{2}$-symmetric F-term from $\mathcal{W}_{\mathcal{H}}$, and
the effective D-term potential
\begin{align}
\label{VD}
V_{D}=&\frac{g^{2}}{8}\sum_{a=1}^{3}
  \left[H_{u}^{\dagger}\sigma^{a}H_{u}-H_{d}\sigma^{a}H_{d}^{\dagger}\right]^{2}
  +\frac{g_{y}^{*2}}{8}
 \left[|H_{u}|^{2}-|H_{d}|^{2}\right]^{2}\\
 \nonumber
  &+\left(\frac{2m^{2}}{2m^{2}+m^{2}_{\hat{W}}}\right)\frac{g^{2}}{8}
  \sum_{\hat{a}=4}^{7}
  \left[\mathcal{H}_{u}^{\dagger}\lambda^{\hat{a}}\mathcal{H}_{u}
   -\mathcal{H}_{d}\lambda^{\hat{a}}\mathcal{H}_{d}^{\dagger}\right]^{2} \\
\nonumber
  &+\left(\frac{2m^{2}}{2m^{2}+m^{2}_{Z^{\prime\prime}}}\right)
     \left\{
     \frac{g^{2}}{8}\left[\mathcal{H}_{u}^{\dagger}\lambda^{8}\mathcal{H}_{u}-
          \mathcal{H}_{d}\lambda^{8}\mathcal{H}_{d}^{\dagger}\right]^{2}
+\frac{g_{x}^{2}}{18}
     \left[|\mathcal{H}_{u}|^{2}-|\mathcal{H}_{d}|^{2}\right]^{2}\right\}
\end{align}
where
\begin{align}
\label{couplings}
\frac{1}{g_{y}^{2}}=
\frac{1}{3g^{2}}+\frac{1}{g_{x}^{2}}\,,\quad
g_{y}^{*2}=\frac{g_{y}^{2}}{1+\left(2m^{2}/m^{2}_{Z^{\prime\prime}}\right)}\,,\quad
m_{\hat{W}}^{2}=g^{2}F^{2}\,,\quad m_{Z^{\prime\prime}}^{2}=\frac{4}{9}(3g^{2}+g_{x}^{2})F^{2}\,.
\end{align}
In the supersymmetric limit $m^{2}/F^{2}\rightarrow0$ one gets the
MSSM D-terms. For the hard breaking, $m^{2}/F^{2}\rightarrow\infty$,
the D-terms corresponding to broken generators do not decouple and
the full $SU(3)_{W}\times U(1)_{x}$ expression is recovered. In this
case the additional (non-MSSM) D-terms induce both quadratic and
quartic terms for the physical Higgs at tree-level, and the
additional explicit mass terms would ruin the double protection
mechanism.

Thus we are forced to take the limit $m^{2}/F^{2}\ll1$, in which
case the global $SU(3)_{2}$ is softly broken by MSSM-like D-terms,
so that the Higgs is actually a pseudo-Goldstone boson. This source
of global symmetry breaking contributes to the tree level quartic
self-interaction of the Higgs $\lambda_{0}=(g^{2}+g_{y}^{2})/8$,
while the leading sources for the Higgs mass terms (as well as for
additional quartics) come from loops of the top/stop sector. This
way the large $\log$ contribution to the Higgs mass parameter is
replaced by a smaller effective one, $\delta
m_{H}^{2}\sim-(3m_{s}^{2}/4\pi^{2})\ln f/m_{s}$, as result of the
double protection of the global symmetry and of supersymmetry which
forbid a $\log\Lambda$ dependence of the cut-off
\cite{Birkedal:2004xi,Chankowski:2004mq,Berezhiani:2005pb}. However,
the contribution to the Higgs quartic is even smaller than in the MSSM
\cite{Bellazzini:2008zy} and a large stop mass
$m_{s}\approx$~TeV is needed. This is especially true when
one includes two loop effects, which  usually lower the effective
Higgs quartic further\cite{Kitano:2006gv}. From the expression
(\ref{VD}) for the D-terms, we see that increasing the soft breaking
parameter $m^{2}$ is even worse because it decreases the effective
$U(1)_{y}$ coupling $g^{*}_{y}<g_{y}$ and also introduces large tree
level corrections to the Higgs mass parameter, since this is
breaking both supersymmetry and the SU(3) global symmetry in a hard
way.

\section{Doping the Super-Little Higgs model}
\label{doping}

We propose a simple way to improve the simplest super-little Higgs
model. We present a mechanism which enhances the Higgs quartic
coupling without introducing additional Higgs mass terms. We have
seen above that in the supersymmetric limit the global symmetry
breaking via D-terms induces only a Higgs quartic coupling, but no
quadratic term. This gives us the hint: if we find a way to increase
the effective low energy gauge coupling $g$ or $g^{*}_{y}$ then we
increase the tree level quartic $\lambda_{0}$ too, keeping the
quadratic tree level Higgs coupling untouched. This can be achieved
by extending the gauge group once more. For example assume that we
have a gauge extension $G_{1}\times G_{2}$ with gauge couplings
$g_{1}$ and $g_{2}$ respectively. Breaking spontaneously to the
diagonal subgroup $G_{1+2}$ at some scale $\Omega$, the low energy
gauge coupling is $1/g^{2}_{1+2}=1/g^{2}_{1}+1/g^{2}_{2}$. However,
as we saw above, the effective D-terms in the low-energy theory are
determined by the same low-energy gauge coupling only in the
supersymmetric limit. If we keep the ratio $m^{2}_{soft}/\Omega^{2}$
finite, where $m_{soft}$ is a soft breaking term of a field charged
only under one of two $G_{i}$, the effective gauge coupling
$g_{eff}$ in front of the D-terms turns out to be slightly bigger,
$g_{eff}\geq g_{1+2}$ \cite{Batra:2003nj,Maloney:2004rc}, achieving
our goal of raising the Higgs quartic couplings.

The simplest way to take advantage of this effect is by adding an
extra U(1)$_{z}$. In order for the U(1)$_z$ to have an effect on the
Higgs couplings, the triplets $\Phi ,\mathcal{H}$ should be charged
under the U(1)$_z$. We also need another field $\Psi$ to break the
additional U(1)$_z$  via its VEV $\langle \Psi \rangle =\Omega$ (the
triplet VEVs would leave an extra unbroken U(1) left over at low
energies). In order for this $\Psi$ not to introduce new global
symmetry breaking terms we will take it to be a singlet under
$SU(3)_{W}\times U(1)_x$. The resulting field content is shown in
Table~(\ref{Higgstable}).
\renewcommand{\arraystretch}{1.25}
\begin{table}[htb]
\begin{center}
\begin{equation*}
\begin{array}{|c|c|c|c|}
\hline
 & SU(3)_{W} & U(1)_{x} & U(1)_{z}\\
 \hline
 \mathcal{H}_{u,d} & 3,\, \bar{3} & \pm1/3 & \pm q' \\
 \hline
 \Phi_{u,d}& 3,\, \bar{3} & \pm1/3 & \pm q \\
 \hline
 \Psi_{u,d} & 1 & 0 & \pm q_{\Psi}\\
 \hline
\end{array}
\renewcommand{\arraystretch}{1}
\end{equation*}
\caption{Charge assignments of the full extended Higgs sector}
\label{Higgstable}
\end{center}
\end{table}

\subsection*{Quartic without quadratic Higgs terms}

In order for the new D-terms to be non-decoupling, the soft breaking
mass $m_\Psi$ for the new field $\Psi$ should be sizeable,
$m_\Psi/\Omega = \mathcal{O}(1)$. The main worry is whether
these non-decoupling effects will only enhance the quartic or also
introduce a large Higgs mass correction reintroducing the fine
tuning. Next we argue that for the case when the U(1)$_z$ charges of
the two triplets are chosen to be equal ($q'=q$) there will be no
quadratic terms introduced. This can be seen as follows. The
low-energy Lagrangian for the light fields should be expressible in
terms of the various D-terms ($D_8,D_y,D_x,D_z$) formed from the light
field $\mathcal{H}$, where
\begin{eqnarray}
&& D_8 = \mathcal{H}^\dagger_u \lambda^8 \mathcal{H}_u-\mathcal{H}_d
\lambda^8 \mathcal{H}_d^\dagger, \nonumber \\
&& D_x = |\mathcal{H}_u|^2 -|\mathcal{H}_d|^2, \nonumber \\
&& D_y= |H_u|^2-|H_d|^2, \nonumber \\
&& D_z=D_x.
\end{eqnarray}
Using the embedding of hypercharge we also find that
\begin{equation}
D_y= \frac{1}{\sqrt{3}} D_8 +\frac{2}{3} D_x .
\end{equation}
So the actual expression of the low-energy Lagrangian for the
scalars should be of the form
\begin{equation}
A D_y^2+ B D_x D_y +C D_x^2.
\end{equation}
$D_x$ does not depend on the Goldstones, however it will have a
non-zero expectation value, so a cross term $D_x D_y$ would indeed
generate a large mass correction to the Higgs. We will show that
this cross term is avoided with the choice $q=q'$.

Since the non-decoupling D-terms from the triplet $\Phi$ give a
large mass to the Higgs, we will take the limit $m/F \to 0$. In this
limit the triplet sector is supersymmetric, and all their effects
should decouple. Similarly, we know that in the supersymmetric limit
$m_\Psi/\Omega \to 0$ for the new field $\Psi$ one should simply
obtain the hypercharge D-term at low energies~\cite{Kawamura:1994ys}, that is $A\to 1,
B,C\to 0$. So one can express these coefficients as
\begin{equation}
A= 1+\frac{m_\Psi^2}{\Omega^2} a(m_\Psi^2/\Omega^2), \ \ B=
\frac{m_\Psi^2}{\Omega^2} b(m_\Psi^2/\Omega^2), \ \ 
C=\frac{m_\Psi^2}{\Omega^2} c(m_\Psi^2/\Omega^2).
\end{equation}
To fix $b,c$ let us now consider another limit, where $m_\Psi/\Omega
\to \infty$. In this case the extra scalar is infinitely heavy, so
it can not have any effect on the low-energy Lagrangian, thus it
should decouple. So the low-energy D-terms should just be the high
energy D-terms with the heavy $\Psi$ eliminated. However, the
$m_\Psi/\Omega \to \infty$ limit has another important effect as
well: since effectively the U(1)$_z$ breaking VEV is set to zero,
one ends up with {\it two} unbroken U(1)'s in this limit: the usual
hypercharge $Y= T^8/\sqrt{3}+ X$ and $Q=T^8/\sqrt{3}+ Z/3q$. So the
D-terms in this limit should be $D_y^2+D_q^2$. For a generic choice
of $q'$ for the Higgs z-charge there will indeed be a contribution
different from $D_y^2$, which will in turn imply the existence of a
large mass term. However, for $q'=q$ we find that $D_q=D_y$, simply
because in this case the two triplets have the same charges, and so
the third components of both triplets should be uncharged under both
unbroken U(1)'s. Thus for $q'=q$ the low-energy Lagrangian in the
$m_\Psi/\Omega \to \infty $ limit is $\propto D_y^2$, which implies that
 $b=c=0$, and so there will be no mass term introduced
for the Higgs (but only a quartic).

\subsection*{The enhanced quartic and tree-level Higgs mass bound}

Let us now explicitly calculate the enhancement of the Higgs quartic
for this doped model. We can take for $\Psi$ a simple superpotential
like $\mathcal{W}_{\Psi}=\kappa_{\Psi}S(\Psi_{u}\Psi_{d}-w^{2})$,
with equal soft masses,
$m_{\Psi_{u}}^{2}=m_{\Psi_{d}}^{2}=m_{\Psi}^{2}>0$, so that
$\langle|\Psi_{u,d}|^{2}\rangle=\Omega^{2}=w^{2}-m_{\Psi}^{2}/\kappa_{\Psi}^{2}>0$.
Integrating out the heavy fields, we are again left at low energy
with the $SU(3)_{2}$ symmetric potential coming from
$\mathcal{W}_{\mathcal{H}}$, and the effective $D$-terms
\begin{equation}
V_{D}=\frac{g^{2}}{8}\sum_{a=1}^{3}
  \left[H_{u}^{\dagger}\sigma^{a}H_{u}-H_{d}\sigma^{a}H_{d}^{\dagger}\right]^{2}
  +\frac{\tilde{g}_{y}^{2}}{8}
 \left[|H_{u}|^{2}-|H_{d}|^{2}\right]^{2}\,.
 \label{Dtermseff}
\end{equation}
Here we neglect contributions of order $m^{2}/F^{2}\ll1$ but we keep
the  ones from $m_{\Psi}^{2}/\Omega^{2}\approx1$ entering in the effective
$U(1)_{y}$ gauge coupling\footnote{The gauge coupling $g_{y}$ is still given by eq.(\ref{couplings}).}
\begin{equation}
\tilde{g}_{y}^{2}=g_{y}^{2}\left\{1+\frac{1}{t_{W}^{2}}\frac{r_{g}^{2}(3-t_{W}^{2})^{2}}{
      \left[1+18r_{g}^{2} r_{q}^{2}\left(g^{2}\Omega^{2}/m_{\Psi}^{2}\right)+r_{g}^{2}(3-t_{W}^{2})\right]}\right\}
\end{equation}
where $t_{W}= g_{y}/g$, $r_{g}=q g_{z}/g$, $r_{q}=q_{\Psi}/3q$. The
main point here is that the Higgs quartic coupling provided by such
doped D-terms is bigger than the MSSM one. In order to see this
explicitly, we parameterize the Higgs triplet fields $H$ (which
determine where the VEVs point) using
  a non-linear sigma model\footnote{We will comment in Section \ref{tuning} about the potential 
  $\mathcal{W}_{\mathcal{H}}$ generating the Higgs VEV $f$.}
\begin{equation}
\mathcal{H}_{u}=\sin\beta\left(H,\sqrt{f^{2}-|H|^{2}}\right)\quad
\mathcal{H}_{d}=\cos\beta\left(H,\sqrt{f^{2}-|H|^{2}}\right)\,.
\end{equation}
Once we plug this parameterization into the full potential, we get
the new tree level upper bound on the Higgs mass $m_{h}$
\begin{equation}
m_{h}^{2}\leq (1-v^{2}/f^{2})\left[m_{Z}^{2}\cos^{2}2\beta+
           m_{W}^{2}\cos^{2}2\beta \frac{r_{g}^{2}(3-t_{W}^{2})^{2}}{
      \left[1+18r_{g}^{2}r_{q}^{2}
      \left(g^{2}\Omega^{2}/m_{\Psi}^{2}\right)+r_{g}^{2}(3-t_{W}^{2})\right]}\right]\,.
\end{equation}
The overall suppression factor $(1-v^{2}/f^{2})$ comes from the
wave-function normalization. We will see in the next section, that
with these enhanced tree-level Higgs mass moderately low stop mass
of few hundred GeV will be allowed, so that the quadratic Higgs
coupling generated at one loop has the natural order of magnitude,
$v^{2}$.
\begin{figure}[htb]
\begin{center}
\includegraphics[scale=0.8]{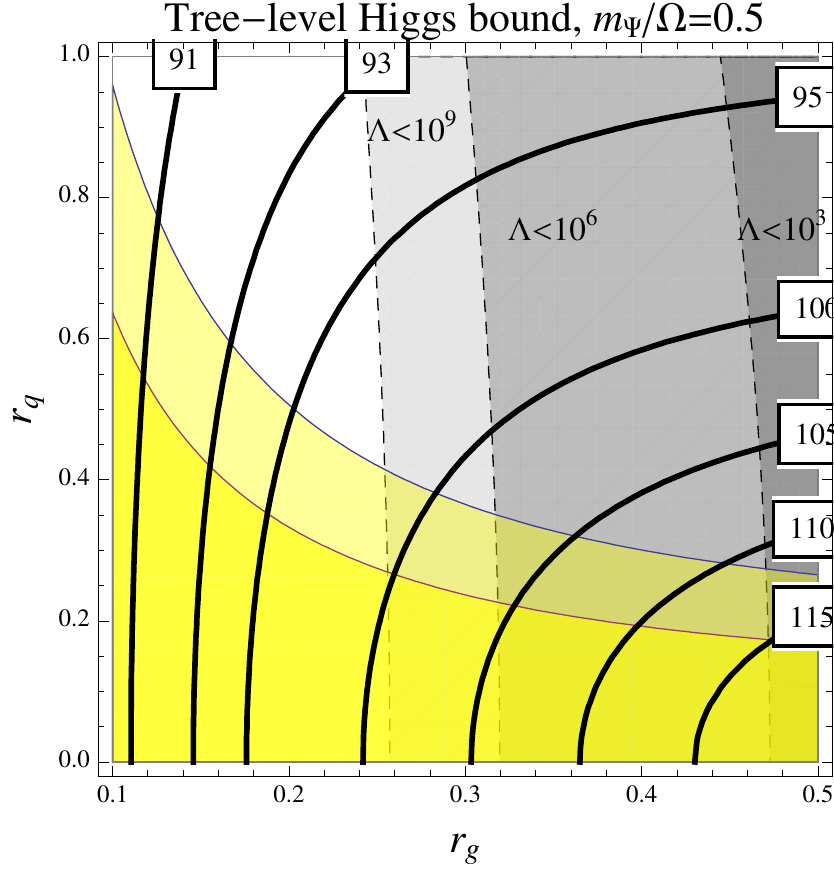}\qquad
\includegraphics[scale=0.8]{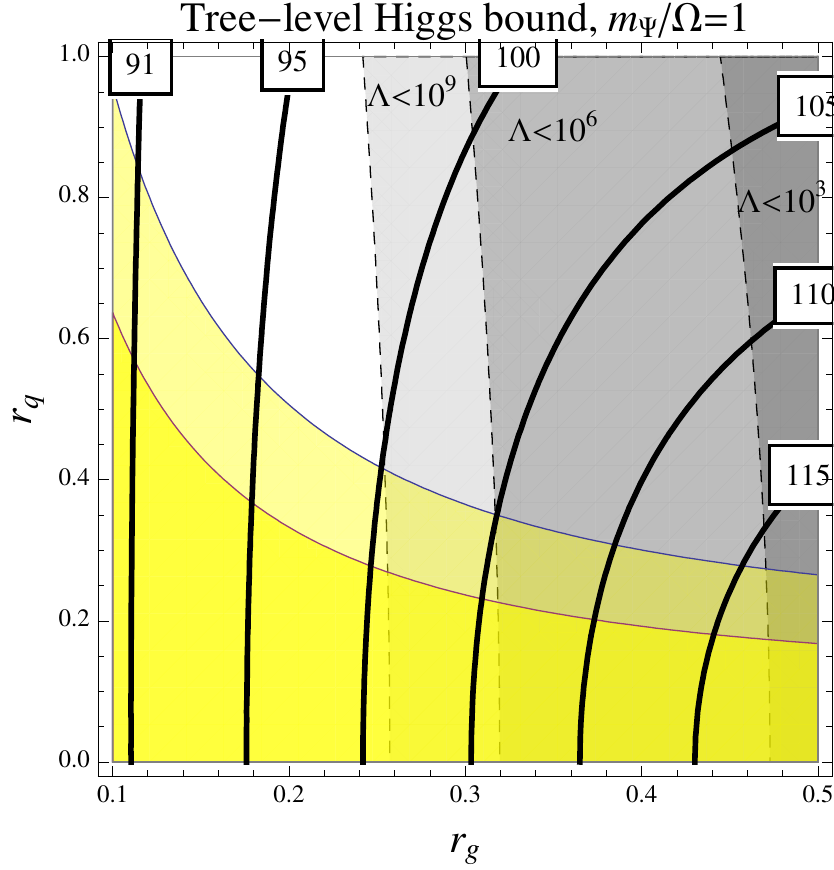}
\caption{Contourplot of the tree level Higgs mass bound (solid black
lines) in the $(r_{g}=qg_{z}/g,\,r_{q}=q_{\Psi}/3q)$ plane. Higgs
masses are expressed in GeV for two different values of the ratio
$m_{\Psi}/\Omega$. Both plots have $\tan\beta=10$. Colored regions
correspond to $m_{Z^{\prime}}$ less than $3$ and  $4.5$ TeV with
$\Omega=12, F=10$ TeV. We also show the regions of parameters
corresponding to Landau poles below $10^3, 10^6$ and $10^9$ TeV, which are shaded gray.}%
\label{fig1}%
\end{center}
\end{figure}
Fig.\ref{fig1} shows the tree level Higgs mass on the plane
$(r_{g},r_{q})$ for fixed $\tan\beta=10$, $F=10$ TeV and $\Omega
=12$ TeV. The two main constraints on the parameter space comes from
the requirement that the $Z^{\prime}$ is sufficiently heavy (to
avoid electroweak precision constraints), and also from ensuring
that the Landau pole of the new U(1)$_z$ is at a sufficiently high
scale. With the fermion matter content presented in the next section
the one-loop expression for the Landau-pole is given by
\begin{equation}
\Lambda_{Landau} \sim M_{Z'} e^{\frac{8\pi^2}{g_z^2 q^2 (138+18
r_q^2)}}. \label{eq:Landaupole}
\end{equation}
In Fig.~\ref{fig1} we show the parameters corresponding to U(1)$_z$
Landau poles of $10^3, 10^6$ and $10^9$ TeV. After imposing the
$Z^{\prime}$ and Landau-pole constraints, the tree-level Higgs mass
is still quite large over a large fraction of the plane (and can
even surpass the LEP bound).

\section{The top sector}
\label{topsect}

In this section we discuss the embedding of the quark and lepton
fields into the theory, focusing ourselves on the radiative
contributions to the Higgs potential from the top/stop loops.

\subsection*{Matter fields and top Yukawa coupling}

Our assignment of quantum numbers, summarized in
Tab.~\ref{mattertable} for the third generation, is the same as in
\cite{SLH}. This choice of charges is anomaly free and generation
universal.
\renewcommand{\arraystretch}{1.25}
\begin{table}[htb]
\begin{center}
\begin{equation*}
\begin{array}{|c|c|c|c|c|c|}
\hline
 & SU(3)_{c} & SU(3)_{W} & U(1)_{x} & U(1)_{z}& U(1)_{em}\\
\hline
Q & 3           & 3 &0&0 & (2/3,-1/3,-1/3) \\
U & \bar{3} & 1 &-2/3& -2 q &-2/3\\
2\times D &  \bar{3} & 1 &+1/3&  +q& 1/3\\
\hline
Q^{\prime} & \bar{3} & 3 & -1/3 & -q & (1/3,-2/3,-2/3)\\
 \bar{Q}^{\prime} & 3 & \bar{3} & +1/3 & +q & (-1/3,+2/3,+2/3)\\
 \hline
 2\times L & 1 & \bar{3} & -1/3 & -q & (-1,0,0)\\
 E & 1 & \bar{3} & +2/3 & +2q & (0,+1,+1)\\
 \hline
\end{array}
\renewcommand{\arraystretch}{1}
\end{equation*}
\caption{The anomaly free and generation independent matter content
that reproduces one generation of quarks and leptons. The presence
of the $Q',\bar{Q}'$ fields allows us to also write down a
renormalizable top Yukawa coupling.} \label{mattertable}
\end{center}
\end{table}
The main difference compared to \cite{SLH} are the Yukawa
couplings.\footnote{In~\cite{SLH} a different representation for the
Higgs sector was used.} The most general superpotential for the top
sector is
\begin{equation}
\mathcal{W}_{top}=m_{Q^{\prime}}\bar{Q}^{\prime}Q^{\prime}+y_{1}\bar{Q}^{\prime}\mathcal{H}_{u}U+
 y_{2}Q \mathcal{H}_{u} Q^{\prime}+\tilde{y}_{1}\bar{Q}^{\prime}\Phi_{u}U+\tilde{y}_{2}Q\Phi_{u}Q^{\prime}\,.
\label{eq:topLag}
\end{equation}
We see the ``collective'' nature of the symmetry breaking pattern:
if only one of the four Yukawa couplings $\tilde{y}_{i}$ or $y_{i}$
is turned on at a time the  global $SU(3)_{1}\times SU(3)_{2}$
symmetry is preserved, while simultaneous non vanishing couplings
$\tilde{y}_{i},\,y_{i}\neq0$, leave only the diagonal symmetry. To
maintain the full global symmetry we will later be working in the
$y_1=\tilde{y}_2=0$ limit.

As long as the  Higgs triplets $\mathcal{H}_{u},\,\Phi_{u}$ have
their VEVs aligned $(0,0,f_{u})$ and $(0,0,F)$ respectively (i.e. no
EWSB occurs), the top stays massless while the heavy top partners
$T_{i=1,2}$ and the heavy bottom partner $B_{1}$ get large masses
\begin{equation}
m_{T_{1}}^{2}=m_{B_{1}}^{2}=m_{Q^{\prime}}^{2}+(y_{2}f_{u}+\tilde{y}_{2}F)^{2}\,,\quad
m_{T_{2}}^{2}=m_{Q^{\prime}}^{2}+(y_{1}f_{u}+\tilde{y}_{1}F)^{2}\,.
\end{equation}
When EWSB occurs via the misalignment of the VEVs,
$\langle\mathcal{H}_{u}\rangle=(0,v_{u},\sqrt{f^{2}_{u}-v_{u}^{2}})$,
the top quark gets a mass proportional to the physical Higgs VEV:
\begin{equation}
m_{t}\equiv y_{t}v_{u}=v_{u}\left(
\frac{F m_{Q^{\prime}}|y_{1}\tilde{y}_{2}-y_{2}\tilde{y}_{1}|}{m_{T_{1}}m_{T_{2}}}\right)\,.
\end{equation}
Note, that the combination $|y_{1}\tilde{y}_{2}-y_{2}\tilde{y}_{1}|$
is simply the determinant of the coupling matrix $(y_{i},\,
\tilde{y}_{i})$, i.e. the two sets of couplings have to be
misaligned as expected by the collective symmetry breaking. Note also,
that the Lagrangian in (\ref{eq:topLag}) would still have a full
SU(3)$\times$SU(3) global symmetry even for the generic choice of
the couplings $y_i,\tilde{y}_i$, except those would be acting on the
linear combinations of the Higgs triplets 
$X=y_1\mathcal{H}_u+\tilde{y}_1 \Phi_u$ and  
$Y=y_2\mathcal{H}_u+\tilde{y}_2 \Phi_u$. In order for the global
symmetries not  to be misaligned with the original global symmetries
we have to choose  $y_{1}=\tilde{y}_{2}=0$ (otherwise the top/stop 
contribution would induce terms like $\propto|X|^{2}\log\Lambda$ 
spoiling the double protection when $y_{1}\neq0$).
Considering the simplest
mass spectrum $m_{T_{1}}\simeq m_{T_{2}}$ with $y_{2}f_{u}\simeq
\tilde{y}_{1}F\simeq m_{Q^{\prime}}$ we get
 the typical sizes of Yukawa couplings: $\tilde{y}_{1}\approx1/5$ and $y_{2}\approx
 2$,
 if we assume that the supersymmetric mass parameter $m_{Q^{\prime}}$
 is of the order of a few TeV.

For the other light SM fermions, we can add the following
superpotential
\begin{equation}
\mathcal{W}=\alpha_{i} Q\mathcal{H}_{d}D_{i}+\tilde{\alpha}_{i}Q\Phi_{d}D_{i}
+\beta_{i} E\mathcal{H}_{d}L_{i}+\tilde{\beta}_{i}E\Phi_{d}L_{i}
\end{equation}
where flavor indices are understood.
In the down quark sector there is another heavy bottom partner $B_{2}$
\begin{equation}
m_{B_{2}}^{2}=(\alpha_{1}f_{d}+\tilde{\alpha}_{1}F)^{2}+(\alpha_{2}f_{d}+\tilde{\alpha}_{2}F)^{2}\,,\quad
m_{b}=v_{d}\left(\frac{Fm_{Q^{\prime}}|\alpha_{1}\tilde{\alpha}_{2}-\alpha_{2}\tilde{\alpha}_{1}|}{m_{B_{1}}m_{B_{2}}}\right)\,.
\end{equation}
We see  that even removing $Q^{\prime}$ and $\bar{Q}^{\prime}$ from the spectrum all 
down quark flavors remain massive. Thus, only the up and charm quarks need  non-renormalizable
operators to get masses. A similar situation occurs for the leptons 
where all the charged states get mass
\begin{equation}
m_{L}^{2}=(\beta_{1}f_{d}+\tilde{\beta}_{1}F)^{2}+(\beta_{2}f_{d}+\tilde{\beta}_{2}F)^{2}\,,\quad
m_{l}=v_{d}\left(\frac{F|\beta_{1}\tilde{\beta}_{2}-\beta_{2}\tilde{\beta}_{1}|}{m_{L}}\right)\,,
\end{equation} 
while the SM neutrinos and the sterile neutrinos $\nu_{s}$ remain massless. 
We will discuss more on $\nu_{s}$ in Section \ref{tuning}.

Gauge coupling unification is not easy to maintain in models based
on SU(3)$_W\times$U(1)$_X$ extensions of the electroweak group.
Interestingly, it was found in~\cite{SLH} that the $\beta$-functions
of the matter content presented here are actually suitable for
one-loop unification into an SU(6) group, except that with the
addition of the vectorlike $Q',\bar{Q}'$ one loses asymptotic
freedom and QCD hits a Landau pole before reaching the GUT scale.
Here we have added one more U(1)$_z$ group, whose Landau pole is
potentially even lower than that of QCD. It remains to be seen
whether a simple extension of these ideas can be made consistent
with perturbative unification.
\begin{figure}[htb]
\begin{center}
\includegraphics[scale=0.8]{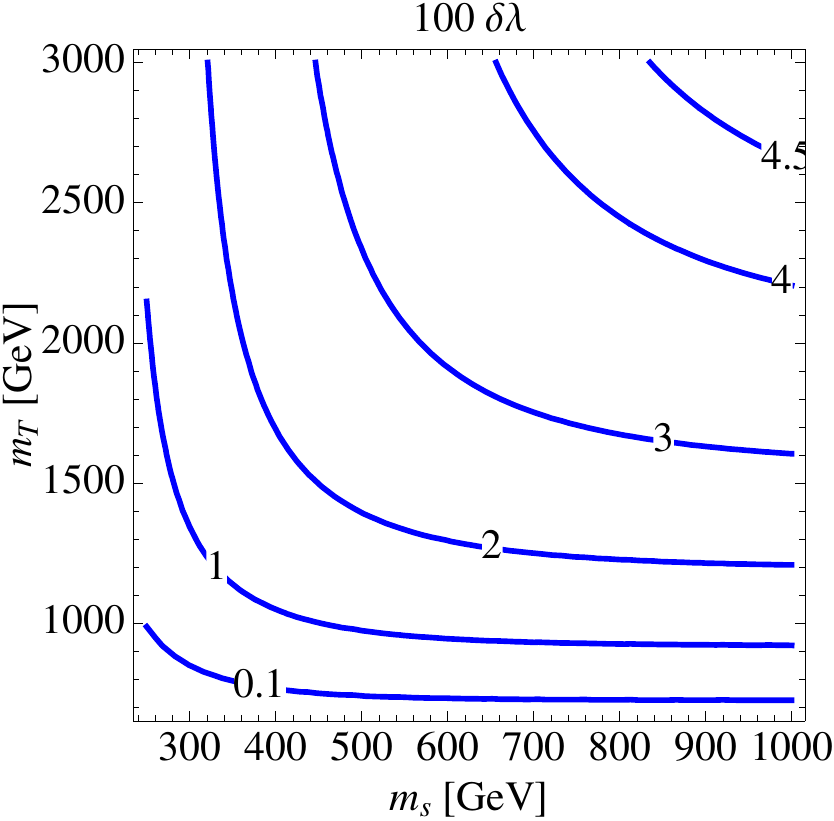}\qquad
\includegraphics[scale=0.8]{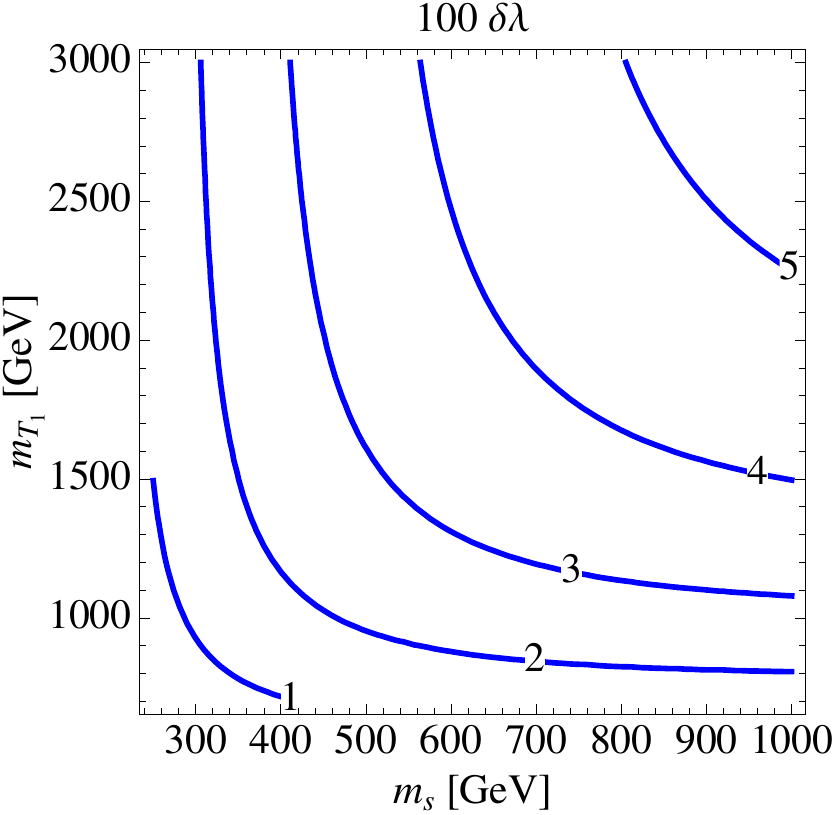}
\caption{Contourplots of $100\times\delta\lambda$ where
$\delta\lambda$ is the radiative correction to the Higgs quartic
coupling in the degenerate limit $m_{T_{1}}=m_{T_{2}}=m_{T}$ (left)
 and in the hierarchical one $m_{T_{1}}\ll m_{T_{2}}$ (right).}%
\label{quartic}%
\end{center}
\end{figure}

\subsection*{Top/stop loop contribution to the Higgs potential}

Next we calculate the radiative potential $\Delta V=\delta
m_{H}^{2}|H|^{2}+\delta\lambda |H|^{4}+\ldots$ generated for the
Higgs by the top/stop loops. Using the Coleman-Weinberg formula
\cite{Coleman:1973jx}, one can explicitly see that $\Delta V$ has no
logarithmic dependence on $\Lambda$, due to the double protection.
For instance, setting for simplicity all soft masses equal,
$m_{s}^{2}\left(|Q^{\prime}|^{2}+|\bar{Q}^{\prime}|^{2}+|U|^{2}+|Q|^{2}\right)$,
 we get
\begin{align}
\label{deltamH}
\delta m_{H}^{2}=-\frac{3y_{t}^{2}\sin^{2}\beta}{8\pi^{2}}
\left(\frac{m_{T_{1}}^{2}m_{T_{2}}^{2}}{m_{T_{2}}^{2}-m_{T_{1}}^{2}}\right)&
\left[\frac{m_{s}^{2}}{m_{T_{1}}^{2}}\ln\left(\frac{m_{T_{1}}^{2}+m_{s}^{2}}{m_{s}^{2}}\right)
-\frac{m_{s}^{2}}{m_{T_{2}}^{2}}\ln\left(\frac{m_{T_{2}}^{2}+m_{s}^{2}}{m_{s}^{2}}\right)\right.\\
\nonumber &\left.
+\ln\left(\frac{m_{T_{2}}^{2}}{m_{T_{1}}^{2}}\right)+\ln\left(\frac{m_{T_{1}}^{2}+m_{s}^{2}}{m_{T_{2}}^{2}
+m_{s}^{2}}\right) \right]+\delta
\end{align}
where $\delta$ is the contribution of the other sectors (like the
gauge/gaugino sector). The top/stop contribution in (\ref{deltamH})
vanishes for $m_{s}\rightarrow0$ or
$m_{T_{i}}\rightarrow0$. The expression of the correction
$\delta\lambda$ to the Higgs quartic coupling is quite long and we
report it in App.\ref{app1}.
\begin{figure}[htb]
\begin{center}
\includegraphics[scale=0.8]{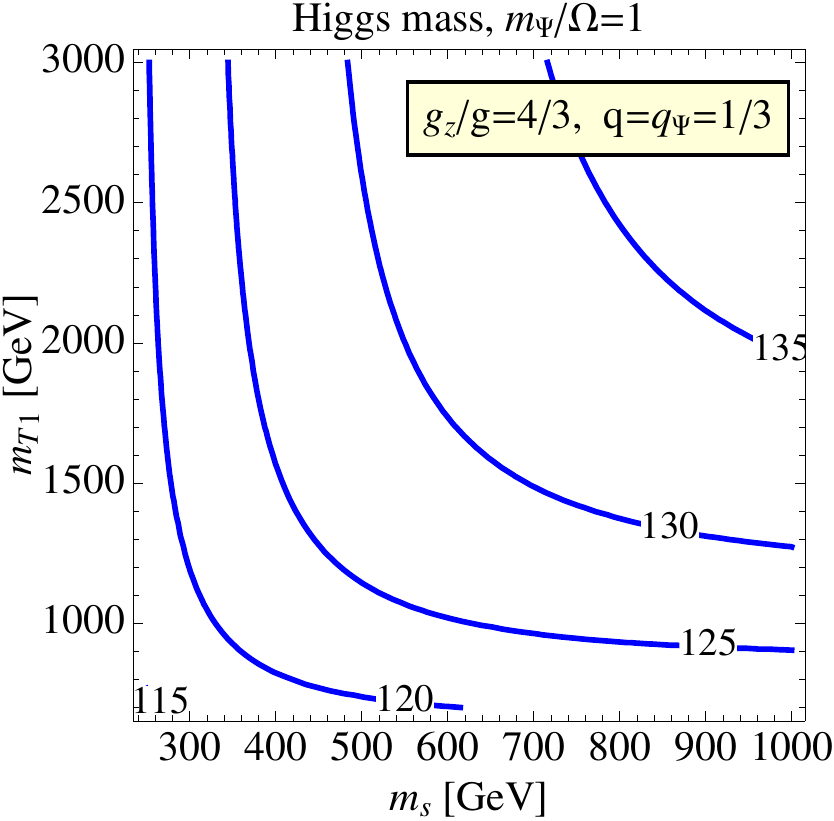}\qquad
\includegraphics[scale=0.8]{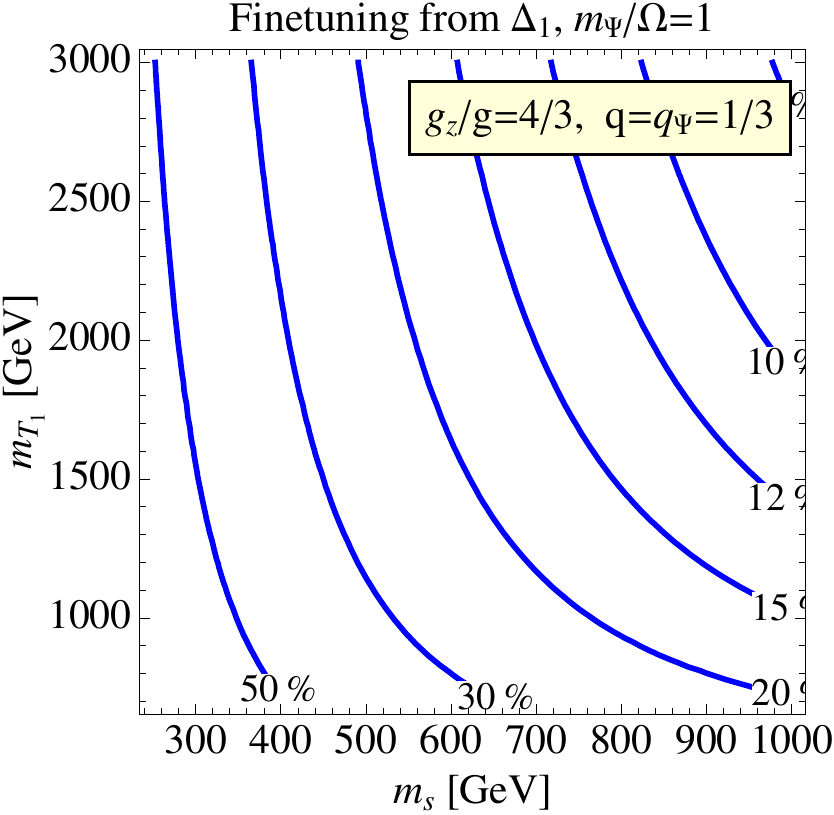}
\caption{Contourplots of the Higgs mass (including radiative
corrections) in GeV (left) and the corresponding fine tuning (right)
for the hierarchical mass spectrum $m_{T_{1}} \ll m_{T_{2}}$ for the
top little-partner and $\tan\beta=10$. The gauge charges for the
U(1)$_z$ correspond to the cutoff scale of $\Lambda =10^3$ TeV.}
\label{mhtune1}%
\end{center}
\end{figure}
The degenerate limit
$m_{T_{1}}=m_{T_{2}}\equiv m_{T}$ is relatively simple:
\begin{equation}
\label{quarticdeg}
\delta\lambda=\frac{3y_{t}^{4}\sin^{4}\beta}{16\pi^{2}}
\left[\ln\left(\frac{m_{s}^{2}}{m_{t}^{2}}\right)
    +\frac{m_{s}^{2}(7m_{s}^{2}+8m_{T}^{2})}{6(m_{s}^{2}+m_{T}^{2})^{2}}
    -\ln\left(\frac{m_{T}^{2}+m_{s}^{2}}{m_{T}^{2}}\right)
    -\frac{4m_{s}^{2}}{m_{T}^{2}}\ln\left(\frac{m_{T}^{2}+m_{s}^{2}}{m_{s}^{2}}\right)
\right]\,.
\end{equation}
The non-log dependent contribution in (\ref{quarticdeg}) is
generated from the interplay of a divergent coefficient
$~(m_{T_{1}}^{2}-m_{T_{2}}^{2})^{-3}$ (in the degenerate limit)
which is in front of a vanishing log in the same limit (see
App.\ref{app2}).

\begin{figure}[htb]
\begin{center}
\includegraphics[scale=0.8]{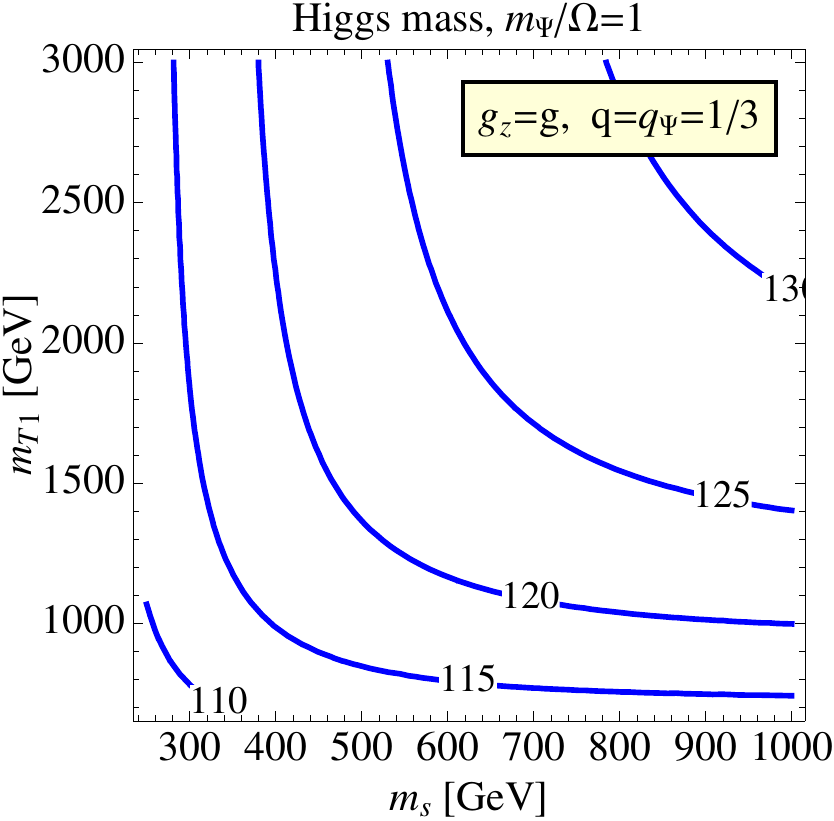}\qquad
\includegraphics[scale=0.8]{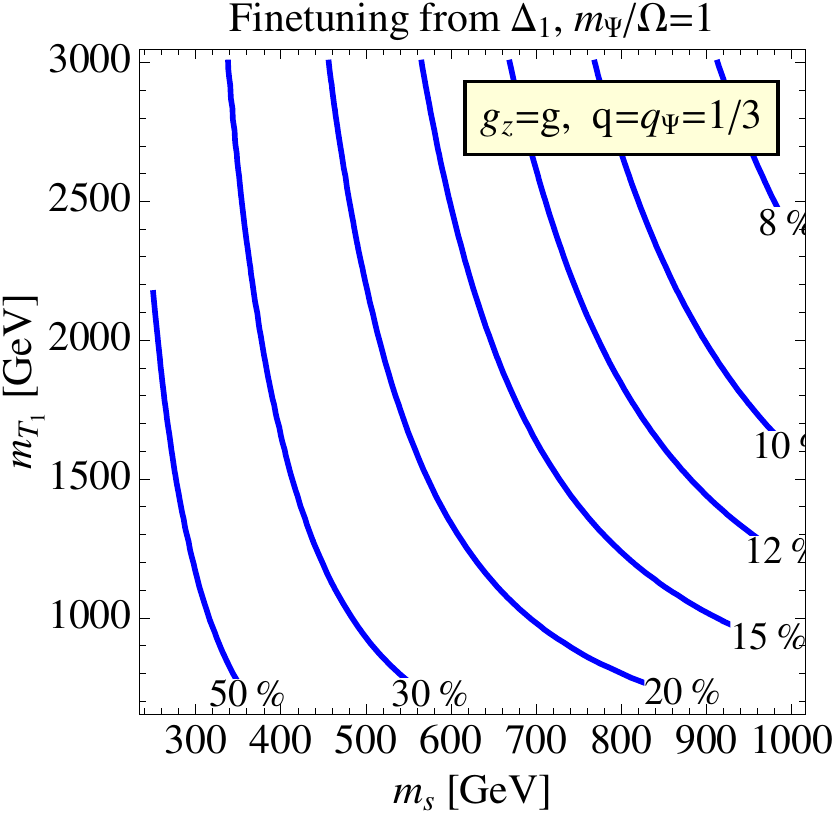}
\caption{The same as in Fig.~\ref{mhtune1} for a different choice of U(1)$_z$ coupling corresponding to a cutoff scale
of $\Lambda =10^6$ TeV.}%
\label{mhtune2}%
\end{center}
\end{figure}
Another simple limit corresponds to taking hierarchical masses for
the little-partners of the top, $m_{T_{1}}\ll m_{T_{2}}$
\begin{equation}
\label{quarticnodeg}
\delta\lambda=\frac{3y_{t}^{4}\sin^{4}\beta}{16\pi^{2}}
\left[\ln\left(\frac{m_{s}^{2}}{m_{t}^{2}}\right)
    +\ln\left(\frac{m_{T_{1}}^2}{m_s^2+m_{T_1}^2}\right)+\frac{2 m^2_s}{m_{T_1}^2}
    \ln\left(\frac{m_s^2}{m_s^2+m_{T_1}^2}\right)
\right]\,.
\end{equation}
A plot of the shift in the quartic due to the top/stop loops for
both limiting cases in shown in Fig.~\ref{quartic}.
 Looking at Figs.~\ref{mhtune1}-\ref{mhtune2}, we see that Higgs
 masses as heavy as $~135$ GeV  are allowed, with less than 10\% fine tuning. The stop can
 be rather light, for example $m_{s}\approx400$ GeV is allowed. In addition, the little-partners
of the top are relatively light, they can be as light as
$m_{T_{1}}\simeq700$ GeV.
\begin{figure}[htb]
\begin{center}
\includegraphics[scale=0.8]{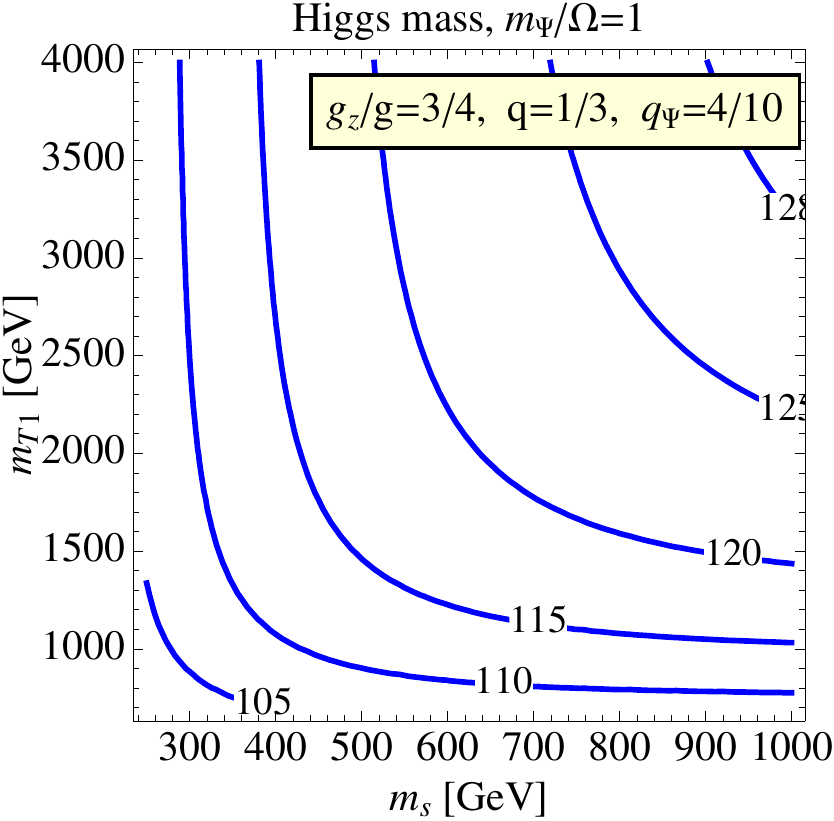}\qquad
\includegraphics[scale=0.8]{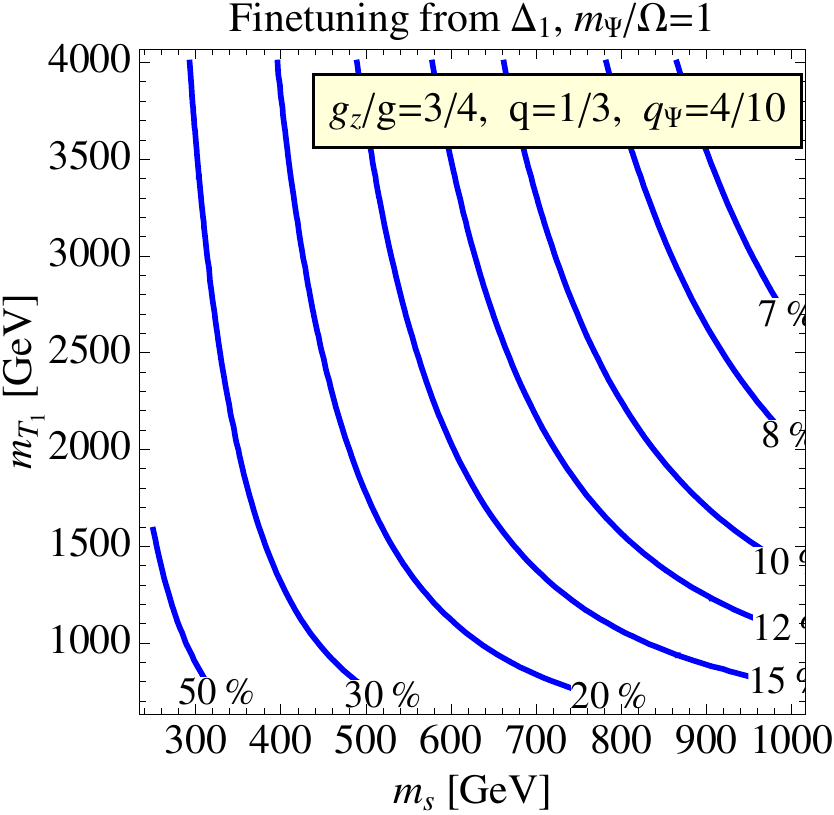}
\caption{The same as in Fig.~\ref{mhtune1} for a choice of U(1)$_z$ coupling corresponding to a high
cutoff scale of $\Lambda =10^9$ TeV. The Higgs mass can still be sufficiently heavy with low
tuning.}%
\label{mhtune3}%
\end{center}
\end{figure}

\section{Tuning and LHC signals}
\label{tuning}

\subsection*{Possible sources of tuning}

Let us now discuss how natural this model is, and how the tuning is
reduced compared to the MSSM and other extensions.
Fig.~\ref{quartic} shows the enhanced quartic coupling due to the
U(1)$_z$ D-term. Fine tuning in the MSSM is usually measured by the
quantity 
\begin{equation}
	\Delta_{1}=|\delta m_{H}^{2}|_{\delta=0}/(m_{h}^{2}/2),
\end{equation}
which shows how much cancellation is needed to compensate the
top/stop loops with the other contributions $\delta$. We are
presenting $\Delta_1$ for three different choices of the charges and
couplings in the U(1)$_z$ gauge sector in
Figs.~\ref{mhtune1}-\ref{mhtune3}  (we actually plot $100/\Delta_1 \%$). The three choices correspond to
cutoff scales (determined by the U(1)$_z$ Landau pole) of $10^3,
10^6$ and $10^9$ TeV. Clearly, the bigger one takes the U(1)$_z$
charges, the bigger  the shift in the Higgs mass, however the Landau
pole will obviously hit earlier. As explained in the Introduction,
in the MSSM there are two origins for the tuning in $\Delta_1$: the
large logarithm from running from a high scale, and the large loop
from a heavy stop needed to raise the Higgs mass. In our case the
stop can be significantly lighter due to the U(1)$_z$ D-term, and
the logarithm is always cut off by the global symmetry breaking
scale $f$. However, if the cutoff scale always turned out to be low
then the double protection would actually not play that important a
role. This is why it is important to emphasize the case in
Fig.~\ref{mhtune3}: here the cutoff is $\Lambda =10^9$ TeV, so the
MSSM logarithm would be of order 20. A gauge extended MSSM with
non-decoupling D-terms but without the double protection would have
a tuning 20 times bigger than in our model (assuming the same cutoff
scale). Thus, the tuning from $\Delta_1$ in this model is
generically highly reduced here.

 However, due to the extended structure there could be other
sources for fine tuning in this model. For example, the soft terms
$m_{\Phi_{u,d}}^{2}$ have to be nearly universal at the scale $F$,
since their difference contributes to the Higgs quadratic coupling
via D-terms. The same argument applies for $m_{\Psi_{u,d}}^{2}$.
Assuming universality at the cutoff, we need to tune by
$\Delta_{2}=(m_{\Phi_{u}}^{2}-m_{\Phi_{d}}^{2})/v^{2}\approx3\tilde{y}_{1}^{2}/(8\pi^{2})m_{s}^{2}/v^{2}\ln\Lambda/F$
and $\Delta_{3}=(m_{\Psi_{u}}^{2}-m_{\Psi_{d}}^{2})/v^{2}\approx
g_{z}^{2}/(8\pi^{2}) S\ln\Lambda/F$ where $S={\rm
Tr}[Z_{i}m_{i}^{2}]$ and $Z_{i}$ is the $U(1)_{z}$ generator. For
typical values of couplings and soft masses, the tuning $\Delta_{2}$
is totally under control resulting in $\Delta_{2}=\mathcal{O}(1)$,
even for $\Lambda\approx10^{15}$ GeV. The source $\Delta_{3}$
identically vanishes if we assume the RG-invariant condition $S=0$
which can be imposed by a $Z_{2}$ symmetry in the $\Psi$ sector.
Even allowing a non-vanishing $S$ as large as
$S\approx(5\,\rm{TeV})^{2}$, $\Delta_{3}$ corresponds to $10\%$ of
finetuning if supersymmetry is broken at a low scale,
$\Lambda\approx100$ TeV, like in gauge mediation. 
Note that the hierarchy among $m_{\Psi}$ and any other soft mass $m_{soft}$ 
it is not a problem because the induced loop corrections  
are suppressed by the charges, the gauge couplings and the loop factor, 
$\delta m_{soft}^{2}/m_{soft}^{2}\sim (m_{\Psi}^{2}/m_{soft}^{2})
(q^{2}q^{2}_{\psi}g_{z}^{4})/(16\pi^{2}) \log\Lambda/f=\mathcal{O}(1)$.

Finally, we comment here about the superpotential $\mathcal{W}_{\mathcal{H}}$ 
for the light triplets.
As long as it is $SU(3)$ symmetric, its specific form does not really matter 
for the low energy physics of the lightest Higgs boson $H$. 
The only important question is whether there is sizable tuning hiding in this sector in order to get the 
 VEV  $f$ much smaller than $F$ at large $\tan\beta$. Indeed 
the effective D-terms (\ref{Dtermseff}) can not provide for a quartic 
coupling when $\mathcal{H}_{u,d}$ point along the third direction of the triplets.
However, a sizable quartic coupling (together  with a negative quadratic driving the breaking) 
for $\mathcal{H}_{u}$ is actually radiatively generated by the Yukawa coupling $y_{2}$. 
The tuning associated to this breaking is milder than $10\%$ if $y_{2}\gtrsim 1.2$ 
(or even less depending on the scale $\Lambda$ where the soft terms are generated).
Such values of $y_{2}$ are small enough to maintain perturbativity up to scales bigger 
 than $\Lambda=10^9$ TeV.

An alternative way to get a sizable quartic for $\mathcal{H}_{u}$ 
is to extend the Higgs sector as in \cite{Bellazzini:2008zy},
 introducing symmetric representations $\mathcal{Z}_{u,d}$ of $SU(3)$
  getting small VEVs $f_{Z}\ll f$. While those fields do not modify the D-terms 
and the tree level Higgs mass formula, they slightly lower the $U(1)_{z}$ Landau pole. 
Keeping the same benchmark values 
 $\Lambda=10^{3},\,10^{6},\,10^{9}$ TeV as above, this means
that we have to choose a slightly smaller $g_{z}$, decreasing
 in fact the tree level Higgs mass by only of $2\div3$ GeV.
\subsection*{Particles within the LHC reach}

\begin{figure}[htb]
\begin{center}
\includegraphics[scale=0.7]{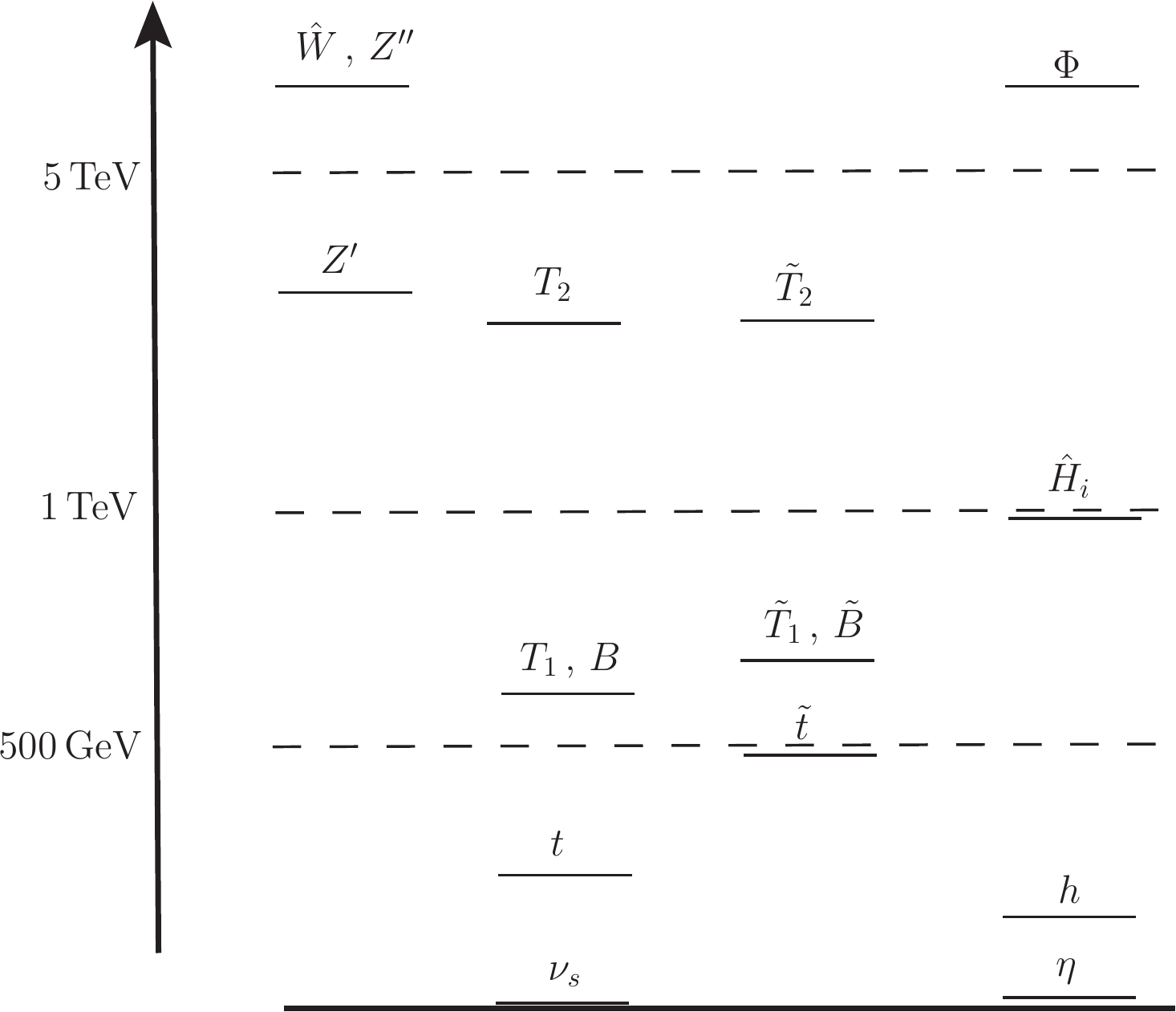}
\caption{A typical mass spectrum for the states beyond the MSSM
assuming the hierarchical pattern for the top partners $m_{T_1}\ll
m_{T_2}$. We do not show most of the MSSM states which can be as
light as 400 GeV.}
\label{spectrum}%
\end{center}
\end{figure}

The model predicts various new particles  potentially within the
reach of the LHC. Here we summarize the main features which are
generically present for a large part of the parameter space. In
Fig.~\ref{spectrum} we sketch a typical mass spectrum focusing only
on the most important properties. The familiar MSSM states can all
be light, with soft breaking masses of the sfermions as low as 400
GeV. We are not including most of these states in
Fig.~\ref{spectrum}, but rather focus on the modes present due to
the gauge extension. A detailed study of the full mass spectrum
varying the input parameters is beyond the scope of this paper.

The enlarged gauge structure leads to the presence of new gauge
bosons. Besides the $W^{\prime}$and $Z^{\prime\prime}$ with masses
of order $\sim g F$ in the multi-TeV range, there is a $Z^{\prime}$
with mass of few TeV, controlled by $\sim q_{\Psi}g_{z}\Omega$. Its
lower bound (from electroweak precision constraints) puts the only
serious constraint on $\Omega\gtrsim5$ TeV.

The best candidate for observing the $SU(3)$ structure is the
fermionic ``little-partner''  of the top, $T_{1}$. Its mass is
directly linked to the global symmetry breaking scale $f$, and can
be as light as $700$ GeV (at which point one may have to start
worrying about loop-level electroweak precision constraints). Since
$T_1$ is even under R-parity (which is assumed throughout the
paper), it can be singly produced at the LHC via $Wb$ fusion or
Drell-Yan, likely decaying into $ht$, $Zt$ or
$Wb$~\cite{Han:2005ru}.

Among the extra fermions there are two $SU(2)_{W}$ sterile neutrinos
$\nu_{s}$ per generation from the third component of $L$. They arise
from the embedding of $SU(2)_{W}$ singlets inside $SU(3)_{W}$
triplets rather than $SU(3)_{W}$ singlets~\cite{SLH}, which is the
only known generation independent charge assignment for the SM
matter content that ensures anomalies cancellation. As a result,
these $\nu_s$'s  are not completely sterile and could in principle
be produced at accelerators. The model can be modified by adding
more singlets to give Dirac masses to the $\nu_s$'s of order
$\sim\mathcal{O}(f)$. Such scalars are already naturally present in
the $SU(6)$ embedding discussed in~\cite{SLH}.

The Higgs sector is very similar to the MSSM in the decoupling limit
and large $\tan\beta$, with $m_{A^{0},\hat{H}}\gtrsim500$ GeV, where
$\hat{H}$ are the radial modes of $\mathcal{H}$. It provides a
natural framework for incorporating the mechanism of~\cite{muBmu}
for eliminating the $\mu-B_\mu$-problem in the context of gauge
mediation. The Higgs can be as heavy as $135$ GeV due to the tree
level enhancement of the quartic governed by the ratio
$m_{\Psi}/\Omega\sim1$ in the effective the D-terms. Thus, a smaller
contribution from the top/stop loop is needed and a quite light stop
around $m_{s}\sim400$ GeV is allowed. Hence the tuning of generic
SUSY theories is drastically reduced here, to levels better than 10\% .

The main difference with respect to the  MSSM Higgs sector is the
presence of an axion-like state $\eta$ coming from a residual global
$U(1)$ symmetry acting on the third components of the triplets
$\mathcal{H}_{u,d}$. This symmetry is explicitly broken by the
Yukawas which will generate a mass at two loops. Alternatively, if
both $y_{1}$ and $\tilde{y}_{1}$ are switched on (but still with one
much smaller than the other to preserve the collective symmetry
breaking), a one-loop mass $\mathcal{O}(1)$ GeV for $\eta$ is
generated, avoiding all the astrophysical constraints.

\section{Conclusions}
\label{conclusions}

We presented an extension of the MSSM which is free of the usual
percent level (or worse) fine tuning of the Higgs mass parameters.
This insensitivity  of the Higgs mass to high scales is achieved via
a double protection (or ``super-little Higgs'') mechanism, whereby
the supersymmetric Higgs is also a pseudo-Goldstone boson of a
global symmetry broken around the TeV scale. While such models are
in principle very appealing, they have usually suffered from the
reduction of the Higgs quartic coupling from the extended sector.
The novelty of our model is to use non-decoupling D-terms in order
to enhance the quartic coupling. For this we enlarge the gauge
structure to incorporate an additional U(1) gauge group, where the
VEV of the field breaking this symmetry is comparable to the SUSY
breaking mass for this field. This will automatically generate an
extra contribution to the Higgs quartic. With an appropriate choice
of the gauge charges one can achieve that the extra contribution to
the Higgs potential does not contain a quadratic term, thus preserving
the double protection mechanism.

The particular model we constructed is an extension of the previous
super-little Higgs models (or supersymmetrized versions of the
simplest little Higgs models) based on the gauge group
$SU(3)_W\times U(1)_x\times U(1)_z$ with an extended global symmetry
$SU(3)\times SU(3)$. This model can be extended to include SM
fermions in anomaly free generation universal representations.

We have carefully analyzed the Higgs potential of this theory,
taking into account the effect of the non-decoupling D-terms and the
one-loop corrections from the top/stop sector. We have found that it
is fairly easy to find regions of the parameter space where the
Higgs mass is well above the LEP bound (and could be as heavy as 135
GeV) with a fine-tuning of less than 10\%. The MSSM superpartners
could be as light as $400$ GeV, while $Z^{\prime}$s dangerous for
electroweak precision corrections can be pushed into the multi-TeV
regime. The Landau pole of the new U(1) symmetry can be separated
from the new physics scales relevant here by several orders of
magnitude. The lightest non-MSSM states that carry SM quantum
numbers are the top partners, some of which could be below 1 TeV
(possibly as light as 700 GeV). Other new light states include a
weakly coupled axion like state (with mass in the GeV range) and
potentially light sterile neutrinos (which are not completely
sterile since they transform under the extended gauge group). It
remains to be seen whether the ideas presented here can be extended
to a model incorporating perturbative unification.

\section*{Acknowledgements}

We thank Riccardo Barbieri, Jay Hubisz, Slava Rychkov and Alvise Varagnolo for
useful discussions. A.D. thanks the Cornell particle theory group
for its hospitality while this work was initiated. The research of
B.B., C.C. and A.W. have been supported in part by the NSF grant
number PHY-0355005 and C.C. was also supported in part by a
US-Israeli BSF grant.

\appendix

\section{The radiative Higgs quartic coupling}
\label{app1}

We report here the full expression for the radiative correction to
the Higgs quartic coupling from the top/stop sector, still assuming
universal soft masses $m_{s}$:
\begin{align}
\delta\lambda=&\frac{3y_{t}^{4}}{16\pi^{2}}\left\{\ln\left(\frac{m_{s}^{2}}{m_{t}^{2}}\right)+
 \left(\frac{2m_{s}^{2}m_{T_{1}}^{6}(m_{T_{1}}^{2}-2m_{T_{2}}^{2})}{m_{T_{1}}^{2}m_{T_{2}}^{2}(m_{T_{1}}^{2}-m_{T_{2}}^{2})^{3}}\right)\ln\left(\frac{m_{s}^{2}}{m_{s}^{2}+m_{T_{2}}^{2}}\right)\right.\\
\nonumber
&\left.+\left(\frac{m_{T_{1}}^{2}m_{T_{2}}^{6}(3m_{T_{1}}^{2}-m_{T_{2}}^{2})}{m_{T_{1}}^{2}m_{T_{2}}^{2}(m_{T_{1}}^{2}-m_{T_{2}}^{2})^{3}}\right)\ln\left(\frac{m_{T_{1}}^{2}}{m_{T_{1}}^{2}+m_{s}^{2}}\right)+\left( m_{T_{1}}^{2}\leftrightarrow m_{T_{2}}^{2}\right)\right\}\,.
\end{align}
The degenerate limit $m_{T_{1}}=m_{T_{2}}=m_{T}$ is finite and is given in formula (\ref{quarticdeg}).
This expression has been derived using the Coleman-Weinberg potential \cite{Coleman:1973jx}
\begin{equation}
\Delta V(H)=\frac{1}{32\pi^{2}}\mathcal{S}\rm{Tr}\left\{M^{4}(|H|)\left(\ln \frac{M^{2}(|H|)}{\Lambda^{2}}-\frac{3}{2}\right)\right\}
=\rm{const.}+\delta m_{H}^{2}|H|^{2}+\delta\lambda |H|^{4}+\ldots
\end{equation}
where the Higgs field $|H|=|H_{u}|/\sin\beta$ measures the
misalignement between the $SU(3)$ breaking VEVs,
$\mathcal{H}_{u}=(0,|H_{u}|,\sqrt{f^{2}_{u}-|H_{u}|^{2}})$. To
determine $\delta\lambda$ one needs the eigenvalues of the matrix
$M$ up to the fourth order in $|H|$. This calculation is reported in
Appendix~\ref{app2}.

\section{Perturbation theory: the fourth order}
\label{app2}
Given a diagonal real matrix $H$, organized in blocks of degenerate eigenvalues $E^{0}_{n}$,
 we determine the approximate eigenvalues of
 $H+\lambda V$ up to the fourth order in $\lambda$
\begin{equation}
 E_{n}=E^{0}_{n}+\lambda \Delta^{(1)}_{n}+ \lambda^{2} \Delta^{(2)}_{n}+\lambda^{3} \Delta^{(3)}_{n}+ \lambda^{4} \Delta^{(4)}_{n}+\ldots
 \end{equation}
 Here $V$ is a Hermitian perturbation matrix
 which is diagonal inside each block of $H$ (completely removing the degeneracy).
 Following any standard quantum mechanics textbook, we get
 \begin{align}
 \Delta^{(1)}_{n}=& V_{nn}\\
 \Delta^{(2)}_{n}=&\sum_{k\neq n}\frac{V_{nk}V_{kn}}{(E^{0}_{n}-E^{0}_{k})}\\
 \Delta^{(3)}_{n}=&\sum_{k,j\neq n}\frac{V_{nk}V_{kj}V_{jn}}{(E^{0}_{n}-E^{0}_{k})(E^{0}_{n}-E^{0}_{j})}-
    V_{nn}\sum_{k\neq n}\frac{V_{nk}V_{kn}}{(E^{0}_{n}-E^{0}_{k})^{2}}\\
 \Delta^{(4)}_{n}=&\sum_{k,j,l\neq n}\frac{V_{nk}V_{kl}V_{lj}V_{jn}}{[(E^{0}_{n}-E^{0}_{k})(E^{0}_{n}-E^{0}_{l})(E^{0}_{n}-E^{0}_{j})]}\\
 \nonumber
 &-V_{nn}\sum_{k,l\neq n}\frac{V_{nk}V_{kl}V_{ln}}{[(E^{0}_{n}-E^{0}_{k})^{2}(E^{0}_{n}-E^{0}_{l})]}-
 V_{nn}\sum_{k,l\neq n}\frac{V_{nk}V_{kl}V_{ln}}{[(E^{0}_{n}-E^{0}_{k})(E^{0}_{n}-E^{0}_{l})^{2}]}\\
 \nonumber
 &+ V_{nn}^{2}\sum_{k\neq n}\frac{V_{nk}V_{kn}}{(E^{0}_{n}-E^{0}_{k})^{3}}
 -\sum_{k\neq n}\frac{V_{nk}V_{kn}}{(E_{n}^{0}-E_{k}^{0})^{2}}\sum_{l\neq n}\frac{V_{nl}V_{ln}}{(E_{n}^{0}-E_{l}^{0})}\,.
 \end{align}
In order to use this result in the calculation of the Coleman-Weinberg potential, we need to rotate into a basis in which
$V=M^{2}(|H|)-M^{2}(0)$ is diagonal in each degenerate block.



\begin{thebibliography}{99}
%
%
\bibitem{Barbieri:2000gf}
  R.~Barbieri and A.~Strumia,
  [arXiv:hep-ph/0007265].
%
\bibitem{ArkaniHamed:2001nc}
  N.~Arkani-Hamed, A.~G.~Cohen and H.~Georgi,
  Phys.\ Lett.\  B {\bf 513}, 232 (2001)
  [arXiv:hep-ph/0105239].
%
 \bibitem{Schmaltz:2005ky}
  M.~Schmaltz and D.~Tucker-Smith,
  Ann.\ Rev.\ Nucl.\ Part.\ Sci.\  {\bf 55}, 229 (2005)
  [arXiv:hep-ph/0502182];
  M.~Perelstein,
  Prog.\ Part.\ Nucl.\ Phys.\  {\bf 58}, 247 (2007)
  [arXiv:hep-ph/0512128].
%
\bibitem{Kaplan:1983fs}
  D.~B.~Kaplan and H.~Georgi,
  Phys.\ Lett.\  B {\bf 136}, 183 (1984);
  D.~B.~Kaplan, H.~Georgi and S.~Dimopoulos,
  Phys.\ Lett.\  B {\bf 136}, 187 (1984).
%
%
%
\bibitem{Agashe:2004rs}
  K.~Agashe, R.~Contino and A.~Pomarol,
  Nucl.\ Phys.\  B {\bf 719}, 165 (2005)
  [arXiv:hep-ph/0412089];
%
%
  G.~F.~Giudice, C.~Grojean, A.~Pomarol and R.~Rattazzi,
  JHEP {\bf 0706}, 045 (2007)
  [arXiv:hep-ph/0703164];
%
%
  R.~Barbieri, B.~Bellazzini, V.~S.~Rychkov and A.~Varagnolo,
  Phys.\ Rev.\  D {\bf 76}, 115008 (2007)
  [arXiv:0706.0432 [hep-ph]].
%
%
\bibitem{Hewett:2002px}
  J.~L.~Hewett, F.~J.~Petriello and T.~G.~Rizzo,
  JHEP {\bf 0310}, 062 (2003)
  [arXiv:hep-ph/0211218].
%
  C.~Cs\'aki, J.~Hubisz, G.~D.~Kribs, P.~Meade and J.~Terning,
  Phys.\ Rev.\  D {\bf 67}, 115002 (2003)
  [arXiv:hep-ph/0211124];
  Phys.\ Rev.\  D {\bf 68}, 035009 (2003)
  [arXiv:hep-ph/0303236];
  %
  Z.~Han and W.~Skiba,
  Phys.\ Rev.\  D {\bf 72}, 035005 (2005)
  [arXiv:hep-ph/0506206];
%
  G.~Marandella, C.~Schappacher and A.~Strumia,
  Phys.\ Rev.\  D {\bf 72}, 035014 (2005)
  [arXiv:hep-ph/0502096].
%
\bibitem{Cheng:2003ju}
  H.~C.~Cheng and I.~Low,
  JHEP {\bf 0309}, 051 (2003)
  [arXiv:hep-ph/0308199];
%
%
  JHEP {\bf 0408}, 061 (2004)
  [arXiv:hep-ph/0405243].
%
%
\bibitem{Hubisz:2005tx}
  J.~Hubisz, P.~Meade, A.~Noble and M.~Perelstein,
  JHEP {\bf 0601}, 135 (2006)
  [arXiv:hep-ph/0506042].
%
%
\bibitem{TparityUVcompl}
C.~Cs\'aki, J.~Heinonen, M.~Perelstein and C.~Spethmann,
  arXiv:0804.0622 [hep-ph].

\bibitem{Birkedal:2004xi}
  A.~Birkedal, Z.~Chacko and M.~K.~Gaillard,
  JHEP {\bf 0410}, 036 (2004)
  [arXiv:hep-ph/0404197].
%
%
\bibitem{Chankowski:2004mq}
  P.~H.~Chankowski, A.~Falkowski, S.~Pokorski and J.~Wagner,
  Phys.\ Lett.\  B {\bf 598}, 252 (2004)
  [arXiv:hep-ph/0407242].
%
%
\bibitem{Berezhiani:2005pb}
  Z.~Berezhiani, P.~H.~Chankowski, A.~Falkowski and S.~Pokorski,
  Phys.\ Rev.\ Lett.\  {\bf 96}, 031801 (2006)
  [arXiv:hep-ph/0509311].
%
%
\bibitem{Roy:2005hg}
  T.~S.~Roy and M.~Schmaltz,
  JHEP {\bf 0601}, 149 (2006)
  [arXiv:hep-ph/0509357].
%
%
\bibitem{SLH}
  C.~Cs\'aki, G.~Marandella, Y.~Shirman and A.~Strumia,
  Phys.\ Rev.\  D {\bf 73}, 035006 (2006)
  [arXiv:hep-ph/0510294].
%
%
\bibitem{Bellazzini:2008zy}
  B.~Bellazzini, S.~Pokorski, V.~S.~Rychkov and A.~Varagnolo,
  JHEP {\bf 0811}, 027 (2008)
  [arXiv:0805.2107 [hep-ph]].
%
%
\bibitem{Batra:2003nj}
  P.~Batra, A.~Delgado, D.~E.~Kaplan and T.~M.~P.~Tait,
  JHEP {\bf 0402}, 043 (2004)
  [arXiv:hep-ph/0309149].
 %
 %
\bibitem{Maloney:2004rc}
  A.~Maloney, A.~Pierce and J.~G.~Wacker,
  JHEP {\bf 0606}, 034 (2006)
  [arXiv:hep-ph/0409127].
%

\bibitem{NMSSM}
P.~Fayet,
  Nucl.\ Phys.\  B {\bf 113}, 135 (1976);
 H.~P.~Nilles, M.~Srednicki and D.~Wyler,
  Phys.\ Lett.\  B {\bf 120}, 346 (1983);
 J.~P.~Derendinger and C.~A.~Savoy,
  Nucl.\ Phys.\  B {\bf 237}, 307 (1984);
 J.~R.~Ellis, J.~F.~Gunion, H.~E.~Haber, L.~Roszkowski and F.~Zwirner,
  Phys.\ Rev.\  D {\bf 39}, 844 (1989).

\bibitem{fathiggs}
R.~Harnik, G.~D.~Kribs, D.~T.~Larson and H.~Murayama,
  Phys.\ Rev.\  D {\bf 70}, 015002 (2004)
  [arXiv:hep-ph/0311349];
S.~Chang, C.~Kilic and R.~Mahbubani,
  Phys.\ Rev.\  D {\bf 71}, 015003 (2005)
  [arXiv:hep-ph/0405267]; 
  A.~Delgado and T.~M.~P.~Tait,
  JHEP {\bf 0507}, 023 (2005)
  [arXiv:hep-ph/0504224]; 
   R.~Barbieri, L.~J.~Hall, Y.~Nomura and V.~S.~Rychkov,
  Phys.\ Rev.\  D {\bf 75}, 035007 (2007)
  [arXiv:hep-ph/0607332].

%
\bibitem{Schmaltz:2004de}
  M.~Schmaltz,
  JHEP {\bf 0408}, 056 (2004)
  [arXiv:hep-ph/0407143];
  D.~E.~Kaplan and M.~Schmaltz,
  JHEP {\bf 0310}, 039 (2003)
  [arXiv:hep-ph/0302049].
%
\bibitem{Kawamura:1994ys}
  Y.~Kawamura, H.~Murayama and M.~Yamaguchi,
  Phys.\ Rev.\  D {\bf 51}, 1337 (1995)
  [arXiv:hep-ph/9406245].
%
\bibitem{Kitano:2006gv}
  R.~Kitano and Y.~Nomura,
  Phys.\ Rev.\  D {\bf 73}, 095004 (2006)
  [arXiv:hep-ph/0602096].
%
%
\bibitem{Coleman:1973jx}
  S.~R.~Coleman and E.~J.~Weinberg,
  Phys.\ Rev.\  D {\bf 7}, 1888 (1973).

\bibitem{Han:2005ru}
  T.~Han, H.~E.~Logan and L.~T.~Wang,
  JHEP {\bf 0601}, 099 (2006)
  [arXiv:hep-ph/0506313].
\bibitem{muBmu}
C.~Cs\'aki, A.~Falkowski, Y.~Nomura and T.~Volansky,
  arXiv:0809.4492 [hep-ph].


\end{thebibliography}
\end{document}